\documentclass[journal]{IEEEtran}
\usepackage{graphicx} 
\usepackage{amsmath} 
\usepackage{amssymb}
\usepackage{amsfonts}
\usepackage{caption}
\usepackage{subcaption}
\usepackage{psfrag}
\usepackage{array}
\usepackage{multirow}
\usepackage{multicol}
\usepackage{mathtools, cuted}
\usepackage{setspace}
\usepackage{longtable}
\usepackage{float}
\usepackage{color}
\usepackage{hyperref}
\usepackage{flushend}

\newcommand{\bm}{\mathbf}
\newcommand{\be}{\begin{equation}}
\newcommand{\ee}{\end{equation}}
\newcommand{\bea}{\begin{eqnarray}}
\newcommand{\eea}{\end{eqnarray}}

\newcommand{\br}{{\bm r}}

\newcommand{\bA}{{\bm A}}
\newcommand{\bI}{{\bm I}}
\newcommand{\bW}{{\bm W}}

\newcommand{\bD}{{\bf D}}
\newcommand{\bC}{{\bf C}}
\newcommand{\bB}{{\bf B}}

\newcommand{\bH}{{\bf H}}

\newcommand{\bd}{{\bf d}}

\newcommand{\bs}{{\bf s}}
\newcommand{\bx}{{\bf x}}

\newcommand{\n}{{\bm n}}

\newcommand{\bb}{\mbox{\boldmath{$b$}}}

\newcommand{\SNR}{\mbox{$\Gamma$}}
\newcommand{\complex}{\mathbb{C}}
\renewcommand{\natural}{\mathbb{N}}

\newcommand{\delaymatrix}{\bm{\Pi}}
\newcommand{\dopplermatrix}{\bm{\Delta}}
\newcommand{\delayindex}{l}
\newcommand{\dopplerindex}{k}

\newcommand{\bL}{\mathbf{L}}

\newcommand{\subcarrier}{\Delta f}
\newcommand{\kron}{\otimes}
\newcommand{\bn}{{\bf n}}
\DeclarePairedDelimiter{\ceil}{\lceil}{\rceil}

\newcommand{\power}{\mathcal{P}}
\newcommand{\NomaPowFrac}{\beta}
\newcommand{\SINRprerxDL}{\Upsilon^{\text{Pre-D}}}
\newcommand{\SINRpostrxDL}{\Upsilon^{\text{Post-D}}}
\newcommand{\SINRrxUL}{\Upsilon^{\text{U}}}
\newcommand{\Exptop}{\mathbb{E}}
\begin{document}
	  \title{Non Orthogonal Multiple Access with Orthogonal Time Frequency Space Signal Transmission }
	\author{Aritra Chatterjee, Vivek Rangamgari, Shashank Tiwari 
         and Suvra Sekhar Das,~\IEEEmembership{Member,~IEEE}
\thanks{Accepted for publication in IEEE Systems Journal. Authors are with G. S. Sanyal School of Telecommunications, Indian Institute of Technology, Kharagpur, India. e-mail: aritrachatterjee@iitkgp.ac.in; rkvivek97@gmail.com; shashankpbh@gmail.com; suvra@gssst.iitkgp.ernet.in.}
}
		
		\maketitle
		
		\begin{abstract}

			  Orthogonal time frequency space (OTFS) is being pursued in recent times as a suitable wireless transmission technology for use in high mobility scenarios. In this work, we propose nonorthogonal multiple access (NOMA) based OTFS which may be called `NOMA-OTFS' system and evaluate its performance from system level and link level perspective. 
			  The challenge lies in the fact that while OTFS transmission technology is known for its resilience to high mobility conditions, while NOMA is known to yield high spectral efficiency in low mobility scenarios in comparison to orthogonal multiple access (OMA). 
			  We present a minimum mean square error (MMSE)- successive interference cancellation (SIC) based receiver for NOMA-OTFS, for which we derive expression for symbol-wise post-processing SINR in order to evaluate system sum spectral efficiency (SE). We develop power allocation schemes to maximize the sum SE in the high-mobility version of NOMA.
			  We further design a realizable codeword level SIC (CWIC) receiver using LDPC codes along with MMSE equalization for evaluating link level performance of such practical NOMA-OTFS system. 
			  The system level and link level performance of the proposed NOMA-OTFS system are compared against benchmark OMA-OTFS, OMA-orthogonal frequency division multiplexing (OMA-OFDM) and NOMA-OFDM schemes.
			  From system-level performance evaluation, we observe interestingly that NOMA-OTFS provides higher system sum SE than OMA-OTFS. When compared to NOMA-OFDM, we find that outage SE of NOMA-OTFS is improved at the cost of decrease in mean SE. Whereas link-level results show that the developed CWIC based NOMA-OTFS receiver performs significantly better than NOMA-OFDM in terms of block error rate (BLER), goodput and throughout. 
		\end{abstract}
	
		\IEEEpeerreviewmaketitle
		\begin{IEEEkeywords}
NOMA, OTFS, SIC, LDPC, power allocation, BLER, spectral efficiency.
\end{IEEEkeywords}
	\section{Introduction}
\subsection{Background and Motivation}
\indent We are experiencing new high mobility scenarios such as high speed railways \cite{Ai_2014, He_2016, Hasegawa_2018}, unmanned aerial vehicle (UAV) communications \cite{Hayat_2016}, vehicle-to-vehicle (V2V) communications \cite{Viriyasitavat_2015, Liang_2017} etc., where providing high quality wireless communication service using existing transmission technologies is a challenge \cite{Wu_2016}. Orthogonal frequency division multiplexing (OFDM) is one of the most successful waveforms used is popular broadband wireless communication systems namely DVB-T, DVB-A, DVB-S \cite{Nee_2000, Prasad_2004}, WiFi \cite{IEEE_80211n_standard}, 4G-LTE \cite{Dahlman_2013}. However it is well known that OFDM suffers from inter carrier interference (ICI) due to high Doppler in such scenarios \cite{Das_2008}. Although in the upcoming 5G new radio (NR), the subcarrier bandwidth of OFDM is made flexible \cite{Das_2018book} to adapt to various channel conditions, yet it is limited due to several other constraints as will be discussed in later sections. On the other hand orthogonal time frequency space~(OTFS)~\cite{Hadani_2018_patent} which places signal constellation in delay-Doppler (De-Do) plane as opposed to time-frequency (T-F) plane, is being explored with enthusiasm by researchers across the globe \cite{Monk_2016, Hadani_2017, Raviteja_2018a, Surabhi_2019a}, as it provides great improvements in performance especially in such new high mobility scenarios. \\
\indent A radio access technology (RAT) comprises of transmission technology and multiple access. In this work we focus on multiple access for OTFS transmission technology. \\
\indent There are two broad class of multiple access (MA) technique namely (i) orthogonal multiple access (OMA) and (ii) non orthogonal multiple access (NOMA). In OMA, resource allocation orthogonality is maintained i.e. one resource unit is allocated to only one user.
With reference to OTFS, two types of OMA-OTFS are reported (i) T-F MA OTFS \cite{Khammammetti_2019}: where users are allocated different T-F resources  and (ii) De-Do MA-OTFS \cite{Hadani_2018_patent}: where users are allocated different D-D resources.\\
\indent In contrast to OMA methodology, NOMA schemes allocate more than one user in one resource unit. Power-domain NOMA (PD-NOMA) schemes realized using superposition coding (SC) at the transmitter along with successive interference cancellation (SIC) at the receiver is known achieve the capacity of Gaussian broadcast channel. PD-NOMA schemes are found to significantly outperform orthogonal multiple access (OMA) as well as code-division NOMA schemes \cite{Le_2018} in terms of sum spectral efficiency (SE) performance \cite{Tse_2005_book, Cover_2012}. \\
\indent Therefore we aim to investigate NOMA-OTFS in this work, which has attracted only limited attention till now. We aim to develop and investigate the NOMA-OTFS and compare its performance against NOMA-OFDM and OMA-OTFS.\\
\indent Important aspects pertaining to the implementation of NOMA are (i) division of total available transmit power at BS among users, (ii) user grouping and T-F resource allocation. Such issues are addressed at length for both downlink and uplink directions in \cite{Saito_2013a, Saito_2013b, Parida_2014, Di_2016, Ali_2016, Nain_2017} and references therein. Such resource allocation is done based on either full or partial channel state information (CSI) at transmitter. An overview on resource allocation and performance analysis of power-domain NOMA systems, is available in \cite{Ding_2017_survey, Dai_2018_NOMAsurvey}.\\
\indent As indicated above, in order to achieve optimal gains, NOMA transmitter must be made aware of the user's instantaneous channel coefficients, which change rapidly in high mobility scenarios as considered in this work. The CSI fed back to transmitter becomes outdated very fast which limits the achievable gain of using T-F domain NOMA.\\ 
\indent Further, because of OFDM's limited capability to handle high Doppler restricts the choice of NOMA-OFDM as a RAT in high mobility scenarios. Since OTFS is resilient to high Doppler in comparison to OFDM, we aim to investigate the use of NOMA with OTFS so that multi user extension of OTFS can be achieved in such high-mobility conditions. Such investigation is expected to pave the path for future research on methods for multi-user spectral efficiency enhancement techniques in high mobility scenarios.
\subsection{Related Works and Contribution}
The interplay of two futuristic technologies namely OTFS and NOMA has attracted the attention of researchers as reported in \cite{Ding_OTFSNOMA_2019, Ding_OTFSNOMA_BF_2019}. In \cite{Ding_OTFSNOMA_2019} use of NOMA with OTFS has been presented in order to serve users with heterogeneous mobility profiles for uplink and downlink. In \cite{Ding_OTFSNOMA_BF_2019}, beamforming aspect of `OTFS assisted NOMA' networks has been explored (in presence of multi-antenna base station) to maximize the low-mobility NOMA users' data rate while maintaining high-mobility OTFS user's target data rate.\\
 In light of the limited state-of-the-art available as indicated above, the major contributions of this work are outlined as follows: 
\begin{itemize}
\item The system model in \cite{Ding_OTFSNOMA_2019, Ding_OTFSNOMA_BF_2019} considers only the user with highest velocity is served in De-Do plane (using OTFS scheme). Whereas, the rest of low-mobility users are served in same T-F plane (using OFDM modulation) which are multiplexed in PD-NOMA. Therefore, the user with high mobility (served with OTFS) does not participate in De-Do NOMA transmission. In this work we propose and develop a holistic framework to obtain the De-Do PD-NOMA-OTFS where multiple high-mobility users are served by OTFS in the same De-Do resource block, which is the first such proposal to the best of our knowledge.
\item The pulse shape used in \cite{Ding_OTFSNOMA_2019, Ding_OTFSNOMA_BF_2019} is considered to be ideal in nature, which is not realizable in practice due to time-frequency uncertainty principle \cite{Raviteja_2018a}. Such ideal assumption simplifies the system equations which yields to block circulant system matrices. In this work we consider realizable time domain rectangular pulse which does not offer such simplification. 
\item Furthermore, in \cite[Sec. VII]{Ding_OTFSNOMA_2019}, NOMA users are allocated fixed power without taking into account their channel condition. Such elementary power allocation restricts NOMA gain. We evaluate the performance of NOMA-OTFS with different dynamic power allocation strategies suitably designed for high-mobility environments.
\item In \cite{Ding_OTFSNOMA_2019}, SE results are obtained using Shannon's expression using ideal SIC at the receiver. In this work we compute post-processing symbol-level SINR for practical ICI canceling MMSE with SIC NOMA receiver, which renders the results more close to reality (in Sec. \ref{subsec:Ana_DL} and \ref{subsec:Ana_UL}). 
\item In \cite{Ding_OTFSNOMA_2019, Ding_OTFSNOMA_BF_2019} arbitrary De-Do channel is considered for performance analysis, whereas we have evaluated performance of the proposed NOMA-OTFS in practical ITU De-Do channel model \cite{itu2135}. The results provide better estimate of such NOMA-OTFS in future realistic scenarios for 5G and beyond.
\item The framework developed in this work is made flexible so as to handle OTFS and OFDM in an unified matrix representation. It is also worth noting that the modified OFDM framework we adopt in this work use block cyclic prefix (CP) along with MMSE equalizer. Accordingly we analyze the performance of OFDM with block processing and ICI canceling receiver for comparison against OTFS. 
\end{itemize}
The above discussions are for system level performance evaluation. While SE performance analysis gives us one perspective, it is also vital to evaluate the link level performance for such NOMA-OTFS system in order to have a comprehensive view of the performance of such newly proposed system namely NOMA-OTFS. Accordingly, the following are included:
\begin{itemize}
 \item In order to get a link level performance estimation, we need to develop a receiver for NOMA-OTFS system. Accordingly we have developed a codeword level low density parity check (LDPC)-SIC receiver which uses symbol-level log-likelihood ratio (LLR) values of the MMSE based ICI canceling receiver (see Sec. \ref{sec:link_level_OTFS_NOMA}), which to the best of the authors' knowledge first such attempt.
 \item The performance of such realistic LDPC enabled MMSE-SIC receiver is further compared with NOMA-OFDM and OMA-OTFS in terms of block error rate (BLER), throughput (in bits/sec/Hz) and goodput (in bits/sec/Hz).
 \item A comprehensive performance analysis taking into account the system level and link level performance has been presented in this work (see Sec. \ref{sec:result_discussion}). 
\end{itemize}

\textit{Notations}: We use the following notations throughout the paper. We let $x$, $\bx$ and $\bm X$ represent scalars, vectors and matrices respectively. The superscripts $(.)^{\rm T}$and $(.)^{\dagger}$ indicate transpose and conjugate transpose  operations, respectively. $\bI_N$ and $\bW_L$ represents identity matrix with order $N$ and $L$-order normalized IDFT matrix respectively. Kronecker product operator is represented by $\kron$. The Frobenius norm of any matrix $\bm X$ is denoted by $||\bm X||_F$. $diag[.]$ denotes a diagonal matrix whose diagonal elements are formed by the elements of the vector inside.
$circ\{.\}$ denotes a circulant matrix whose first column is given by the vector inside. The expectation parameter is denoted by $\mathbb{E}[.]$. Column-wise vectorization of matrix $(.)$ is represented by $vec\{.\}$.  
The ceiling operator is denoted as $\ceil{.}$. 
\textcolor{black}{$\natural[a~~b]$ represents the set of natural numbers ranging from $a$ to $b$}.  $j=\sqrt{-1}$.  \\
\section{OTFS Signal Model}\label{sec:OTFS_basic}
We consider a multi-carrier and multi time-slot system with total $T_f$ sec. duration and $B$ Hz. bandwidth. We have total $M$ number of sub-carriers having $\subcarrier$ sub-carrier bandwidth and $N$ number of symbols having $T$ symbol duration, thus $B=M\subcarrier$ and $T_f=NT$. \\
\indent For a user (termed as $i$-th user henceforth), the QAM modulated Delay-Doppler data symbols, $d_{i}(k,l)\in \complex$, $k\in \natural[0~N-1]$, $l\in \natural[0~M-1]$, are arranged over Doppler-delay lattice $\Lambda=\{(\frac{k}{NT},~\frac{l}{M \subcarrier})\}$. Data symbols $d_i(k,l)$ is mapped to time-frequency domain data $X_i(n,m)$  on lattice $\Lambda^{\perp}=\{(nT,~m\subcarrier)\}$, $n\in \natural[0~N-1]$ and $m\in \natural [0~M-1]$ by using inverse symplectic fast Fourier transform (ISFFT). Thus $X_i(n,m)$ can be given as \cite{Hadani_2017},
\begin{equation}
X_i(n,m)=\frac{1}{\sqrt{NM}} \sum_{k=0}^{N-1}{\sum_{m=0}^{M-1}{d_i(k,l) e^{j2\pi [\frac{nk}{N}-\frac{ml}{M}]}}}.
\end{equation}
  Next, a time-frequency modulator modulates $X_i(n,m)$ to time domain using Heisenberg transform as,
  
\begin{equation}
s_i(t)= \sum_{n=0}^{N-1}{\sum_{m=0}^{M-1}}{( \sqrt{\power}X_i(n,m) )g(t-nT) e^{j2\pi m \subcarrier (t-nT)}},
\end{equation}
where, $g(t)$ is transmitter pulse of duration $T$ and transmit power is denoted by $\power$. Further, $s_i(t)$ is sampled at the sampling interval of $\frac{T}{M}$. We collect samples of $s_i(t)$ in $\bs_i=[s_i(0)~s_i(1) \cdots s_i(MN-1)]$. The QAM symbols $d_i(k,l)$ are arranged in \textcolor{black}{$M\times N$} matrix as,
\begin{equation}
\bD_i= \small{\begin{bmatrix}
	d_i(0,0) & d_i(1,0) & \cdots & d_i(N-1,0) \\
	d_i(0,1) & d_i(1,1) & \cdots & d_i(N-1,1) \\
	\vdots & \vdots & \ddots & \vdots  \\
	d_i(M-1,0) & d_i(M-1,1) & \cdots & d_i(N-1,M-1) \\
	\end{bmatrix}}
\end{equation}
The transmitted signal can be written as matrix-vector multiplication as:
\begin{equation}
\label{eqn:OTFS_Tx_symbols}
\bs_i= \bA {\sqrt{\power}\bd_i},
\end{equation}
where, $\bd_i=vec\{\bD_i\}$.
Finally, $\bA_{\textcolor{black}{MN\times MN}}= \bW_N \kron \bI_M $ denotes the OTFS modulation matrix. A cyclic prefix (CP) of length $L_{CP}'\geq L_{CP}-1$ is appended at the starting of the $\bs$, where $L_{CP}$ is the channel's maximum excess delay length. In order to implement OFDM in the same framework, the modulation matrix is modified as $\bA= \bI_N \kron \bW_M$. %
\section{Delay-Doppler Power-Domain NOMA-OTFS}
In this section, we further extend the OTFS signal model presented in Sec. \ref{sec:OTFS_basic} in order to develop the multi-user De-Do PD-NOMA-OTFS signal models and derive the SINR and SE expressions of the same for both downlink and uplink. We consider $K$ users with high velocity are multiplexed in power domain all of which are served by OTFS (unlike \cite{Ding_OTFSNOMA_2019, Ding_OTFSNOMA_BF_2019}) in both downlink and uplink transmission.
\subsection{De-Do PD-NOMA-OTFS Downlink}
\subsubsection{Transmit Signal Model}\label{subsec:NOMA_OTFS_DL_SigMod}
Among the $K$ high mobility users multiplexed in power domain, we assume The $i$-th user is allocated $\NomaPowFrac_i$ fraction of total power $\power$.  Clearly, $\sum_{i=1}^K \NomaPowFrac_i = 1$. Choice of $\NomaPowFrac_i$ is described in detail in Sec. \ref{sec:Pow_allocate_DL_NOMA}. Following the principle of superposition, the composite transmitted signal from the transmitter intended for all users can be written by modifying \eqref{eqn:OTFS_Tx_symbols} as:
\begin{equation}
\label{eqn:Superposed_Tx_symbols}
\bs= \bA \sum_{i=1}^{K}{\sqrt{\NomaPowFrac_i \power}\bd_i}.
\end{equation}
\indent We consider linear time varying (LTV) channels for all the users. Let, the $i$-th user's channel consists of $P_i$ paths with $h_{p^{i}}$ complex attenuations, $\tau_{p^{i}}$ delays and $\nu_{p^{i}}$ Doppler values for $p^{i}$th path where $p^{i}\in \natural[1~P_i]$. Thus, Delay-Doppler channel spreading function for the $i$-th user can be given as,
\begin{equation}
h_i(\tau,\nu)=\sum_{p^{i}=1}^{P_i}{h_{p^{i}} \delta(\tau-\tau_{p^{i}}) \delta(\nu-\nu_{p^{i}})}, ~ i=1,\cdots, K.
\end{equation}
The delay and Doppler values for $p^{i}$th path is given as $\tau_{p^{i}}=\frac{\delayindex_p^{i}}{M\subcarrier}$ and $\nu_{p^{i}}=\frac{\dopplerindex_p^{i}}{NT}$, where $\delayindex_p^{i}\in \natural[0~ M-1]$ and $\dopplerindex_p^{i}\in\natural[0~N-1]$ are the number of delay and Doppler bins on the Doppler-delay lattice for $p^i$th path.  We assume that $N$ and $M$ are sufficiently large so that there is no effect of fractional delay and Doppler on the performance. We also assume the perfect knowledge of $(h_{p^{i}},~\delayindex_{p^{i}},~\dopplerindex_{p^{i}})$, $p^{i}\in \natural[0~P_i-1]$, at the receiver of $i$-th user, as previously considered in \cite{Hadani_2017, Surabhi_2019a}.
One work on such estimation is given in \cite{Raviteja_2018b}. Let $\tau_{max}^{i}$ and $\nu_{max}^{i}$ be the maximum delay and Doppler spread for users. Channel delay length $\alpha^{i}= \ceil{\tau^{i}_{max}M\subcarrier}$ and channel Doppler length, $\beta^{i}= \ceil{\nu_{max}^{i} NT}$. $L_{CP}=\max\limits_{i=1,\cdots,K}(\alpha^{i})$.\\
\indent At the $i$-th user's receiver, after removal of CP, the received signal can be written as \cite{Raviteja_2018a},
\begin{equation}\label{eqn:Rcvd_DL_OTFS}
\br_i=\bH_i \bs +\bn_i, ~i=1,\cdots, K.
\end{equation}
where, $\bn_i$ is white Gaussian noise vector of length $MN$ with elemental variance $\sigma_\n^2$ and $\bH_i$ is a  \textcolor{black}{$MN\times MN$} channel matrix for $i^{\rm th}$ user which is given by,
\begin{equation}
\label{eq:H_chan_matrix}
\bH_i=\sum_{p^{i}=1}^{P_i}{h_p^{i} \delaymatrix^{\delayindex_p^{i}} \dopplermatrix^{\dopplerindex_p^{i}}}, ~ i=1,\cdots, K,
\end{equation}
with $\delaymatrix_{\textcolor{black}{MN\times MN}}=circ\{[0~1~0 \cdots 0]^T_{MN\times 1}\}$ is a circulant delay matrix and $\dopplermatrix=diag[1~ e^{j2\pi\frac{1}{MN}}~ \cdots e^{j2\pi\frac{MN-1}{MN}}]$ is a diagonal Doppler matrix. Using the above mentioned signal model for the De-Do PD-NOMA-OTFS in downlink, we proceed to evaluate the corresponding SINR and SE experienced by the non-orthogonally multiplexed users. \vspace{-0.5cm}
\subsubsection{Receiver Processing, SINR and SE Analysis}\label{subsec:Ana_DL}
In OTFS, when the signal passes through the LTV channel, it encounters both ICI and inter-symbol interference (ISI), thereby degrading its performance. In order to negate such induced ICI and ISI, we equalize the received signal through LMMSE equalizer, \textcolor{black}{as done in \cite{Tiwari_2019}}. Furthermore, in the later stage, SIC has been performed in order to mitigate the NOMA interference at the receiver, which has been discussed subsequently.\\  
\indent The total effective noise at the $i$-th receiver amounts to:
\begin{equation}
 \label{eq:DL_effective_noise}
 \tilde{\bn}_{i_{DL}} = \sum_{i'=1, i'\neq i}^K \sqrt{\NomaPowFrac_{i'}\power}\bH_{i}\bA\bd_{i'}+\n_i.
\end{equation}
Assuming the total effective noise following Gaussian distribution, LMMSE equalization on the received signal $\br_i$ in \eqref{eqn:Rcvd_DL_OTFS} results in estimated data vector for $i$-th user as given in \eqref{eqn:data_DLMMSEOTFS}, where $\SNR_i$ denotes the average SNR of $i$-th user.
\begin{figure*}
\footnotesize
\begin{align}
\label{eqn:data_DLMMSEOTFS}
\hat{\bd_i}= \sqrt{\NomaPowFrac_i}(\bH_i\bA)^{\dagger}\bigg[\NomaPowFrac_i(\bH_i \bA)(\bH_i \bA)^{\dagger}+ 
\sum_{i'=1, i'\neq i}^K\NomaPowFrac_{i'}(\bH_i \bA)(\bH_i \bA)^{\dagger}+ \frac{1}{\SNR_i} \bI\bigg]^{-1} \br_i
=&\sqrt{\NomaPowFrac_i}(\bH_i\bA)^{\dagger} [(\bH_i \bA)(\bH_i \bA)^{\dagger}+ \frac{1}{\SNR_i} \bI]^{-1} \br_i, 
\end{align} 
\rule{0.97\textwidth}{0.5pt}
\end{figure*}
Rewriting \eqref{eqn:data_DLMMSEOTFS} by using \eqref{eqn:Rcvd_DL_OTFS} and \eqref{eqn:Superposed_Tx_symbols}, we obtain: 
\begin{align}
\label{eqn:MMSE_OTFS_desired_interf_DL}
\hat{\bd_i}= \underbrace{\bB_i\sqrt{(\NomaPowFrac_i\power)}\bd_i}_\text{desired signal}&+\underbrace{\sum_{i'=1, i'\neq i}^K\bB_i\sqrt{(\NomaPowFrac_{i'}\power)}\bd_{i'}}_\text{NOMA interference}\\ \nonumber
&+\underbrace{\bC_i\bn_i}_\text{noise component},~ i=1,\cdots, K,
\end{align}
where, for notational simplicity, we assign $\bC_i = \sqrt{\NomaPowFrac_i}(\bH_i\bA)^{\dagger} [(\bH_i \bA)(\bH_i \bA)^{\dagger}+ \frac{1}{\SNR_i} \bI]^{-1}$ and $\bB_i = \bC_i\bH_i\bA$. At this point, without loss of generality, we consider that from the transmitting BS the distance of $i$-th user is higher than the $(i+1)$-th user for $i=1,\cdots, (K-1)$, thus in terms of received average SNR, it can be written as: $\SNR_1 < \SNR_2 <\cdots< \SNR_{i-1}<\SNR_i<\cdots<\SNR_K$. Thus, we assume that following the principle of NOMA, the $i$-th user will not face any interference due to the signals intended for 1st, 2nd,$\cdots,(i-1)$-th users through perfect SIC\footnote{Consideration of imperfect SIC and subsequent error propagation can be seen as a potential future work \cite{Mahady_2019, Wang_2019}.}.  
Using these assumptions and expanding \eqref{eqn:MMSE_OTFS_desired_interf_DL}, the symbol-wise pre- and post SIC received SINR at any user can be formulated. 
For the $i$-th user, the downlink pre- and post-SIC SINR for $j$-th symbol (denoted as $\SINRprerxDL_{ij}$ and $\SINRpostrxDL_{ij}$ respectively) can be given by \eqref{eqn:PreSIC_SINR_DL} and \eqref{eqn:PostSIC_SINR_DL} respectively at the top of the next page, 
with $i=1,\cdots, K$ and $j=1,\cdots, MN$. $b^i_{pq}$ and $c^i_{pq}$ denote the $(p,q)^{\text{th}}$ elements of $\bB_i$ and $\bC_i$ respectively.
\begin{figure*}
\footnotesize
\begin{minipage}{.5\linewidth}
\begin{equation}
  \label{eqn:PreSIC_SINR_DL}
\SINRprerxDL_{ij} = \frac{\underbrace{\NomaPowFrac_i\power|b^i_{jj}|^2}_\text{desired power}}{ \bigg[\underbrace{\NomaPowFrac_i\power\sum_{l=1, l\neq j}^{MN}|b^i_{jl}|^2}_\text{inter-symbol interference}+ 
\underbrace{\sum_{i'=1, i'\neq i}^{MN}\NomaPowFrac_{i'}\power(\sum_{l=1}^{MN}|b^i_{jl}|^2)}_\text{NOMA interference}   + 
\underbrace{\sum_{l=1}^{MN}|c^i_{jl}|^2\sigma_n^2}_\text{noise power}\bigg]},
\end{equation}
\end{minipage}%
\begin{minipage}{.5\linewidth}
\begin{equation}
  \label{eqn:PostSIC_SINR_DL}
~~~~\SINRpostrxDL_{ij} = \frac{\underbrace{\NomaPowFrac_i\power|b^i_{jj}|^2}_\text{desired power}}{ \bigg[\underbrace{\NomaPowFrac_i\power\sum_{l=1, l\neq j}^{MN}|b^i_{jl}|^2}_\text{inter-symbol interference}+
\underbrace{\sum_{i'=i+1}^{MN}\NomaPowFrac_{i'}\power(\sum_{l=1}^{MN}|b^i_{jl}|^2)}_\text{NOMA interference}   + 
\underbrace{\sum_{l=1}^{MN}|c^i_{jl}|^2\sigma_n^2}_\text{noise power}\bigg]},
\end{equation}
\end{minipage}
\rule{0.97\textwidth}{0.5pt}\vspace{-0.6cm}
\end{figure*}
In OTFS, the SINR achieved in all symbols are nearly same for large $M$ and $N$ values \cite{Hadani_2018b} and thus, henceforth we drop subscript $j$ and represent the pre- and post-SIC SINR of $i$-th user as $\SINRprerxDL_i$ and $\SINRpostrxDL_i$ respectively. 
Thus, the downlink sum rate of the system in bps/Hz is given by: 
\begin{equation}
\label{eqn:DL_sum_rate_expression}
R_{sum}^{DL} = \sum_{i=1}^K \text{log}_2(1+\SINRpostrxDL_i). 
\end{equation}
It is noteworthy that the SE performance presented here for downlink (and subsequently for uplink in Sec \ref{subsec:Ana_UL}) are done for such realizable MMSE-SIC receiver only. SE calculation using log-determinant method\footnote{as usually done in conventional point-to-point multiple-input multiple-output (MIMO) systems \cite{paulraj_book2008}} of the delay-Doppler channel $\bH_i$ is beyond the scope of the current work.
\subsection{De-Do PD-NOMA-OTFS Uplink}
\subsubsection{Transmit Signal Model}\label{subsec:NOMA_OTFS_UL_SigMod}
For uplink OTFS-NOMA, all the $K$ users transmit data simultaneously to the base station in delay-Doppler plane, thus  making it a multiple-access channel (MAC). 
For the sake of simplicity, perfect carrier and clock synchronization among the transmitting users has been assumed. It has also been assumed that both the users are implementing same OTFS grid size ($M, N$). It has also been assumed that the receiver BS has perfect knowledge about the channels from transmitting users. 
The OTFS modulated transmitted vector from $i$-th user is given by: 
\begin{equation}
\label{eqn:Tx_symbols_UL_OTFS}
\bs_{i}^u = \bA\sqrt{\power_i^u}\bd_i^u,
\end{equation}
where $\power_i^u$ and $\bd_i^u$ denote the transmit power and vectorized transmit data of the $i$-th user respectively. The uplink average SNR of the $i$-th user is given by $\SNR_i^u = \power_i^u/\sigma_n^2$. The aggregate received signal at the base station after removal of CP is given by: 
\begin{equation}
\label{eqn:Rcvd_UL_OTFS}
\br_{u} = \sum_{i=1}^K \bH_i^u\bs_i^u + \bn,
\end{equation}
where $\bH_i^u$ denotes the $MN\times MN$ delay-Doppler uplink channel matrix from $i$-th user to the BS. Similar to the downlink scenario presented before, we further proceed to evaluate the SINR and SE experienced by the PD-NOMA-OTFS users in uplink direction in the following section.  
\subsubsection{Receiver Processing, SINR and SE Analysis}\label{subsec:Ana_UL}
During uplink transmission in NOMA, the signal from the users with higher SNR are sequentially decoded and successively canceled from the aggregate signal. For the same user ordering as considered in downlink transmission ($\SNR_1^u<\cdots<\SNR_K^u$), while decoding the $i$-th user's signal, the BS will consider the 1st, 2nd, $\cdots, (i-1)$-th users' signals as noise. Thus, for $i$-th user, the effective noise can be denoted as: 
\begin{equation}
 \label{eqn:UL_noise_ithuser}
 \tilde{\bn}_i = \sum_{i'=1}^{i-1} \sqrt{\power_{i'}^u}\bH_{i'}^u\bA\bd_{i'}^u + \bn.
\end{equation}
Subsequently, the noise variance for the $i$-th user is given by: 
\begin{equation}
 \tilde{\sigma}_{n_i}^2 = \Exptop[\tilde{\bn}_i\tilde{\bn}_i^H] = \sum_{i'=1}^{i-1} \power_{i'}^u\bH_i\bH_i^H + \sigma_n^2\bI.
\end{equation}
After processing the received signal ($\br_u$) through LMMSE equalizer (similar to \eqref{eqn:data_DLMMSEOTFS} for downlink), the estimated data vector for the $i$-th user at the BS can be expressed by \eqref{eqn:data_ULMMSEOTFS}.
\begin{figure*}
\footnotesize
 \begin{align}
\label{eqn:data_ULMMSEOTFS}
\hat{\bd_i^u}&=(\bH_i^u\bA)^{\dagger} [(\bH_i^u \bA)(\bH_i^u \bA)^{\dagger}+
\sum_{i'=1}^{i-1}\frac{\power_{i'}^u}{\power_i^u}(\bH_i^u \bA)(\bH_i^u \bA)^{\dagger}+\frac{\sigma_n^2}{\power_i^u} \bI]^{-1} \br_i
=(\bH_i^u\bA)^{\dagger} [\bH_i^u {\bH_i^u}^{\dagger}+\sum_{i'=1}^{i-1}\frac{\SNR_{i'}^u}{\SNR_i^u}\bH_{i'}^u {\bH_{i'}^u}^{\dagger}+\frac{1}{\SNR_i^u} \bI]^{-1} \br_i,
\end{align}
\rule{0.97\textwidth}{0.5pt}\vspace{-0.7cm}
\end{figure*}
For notational simplicity, we denote $\bC_i^u = (\bH_i^u\bA)^{\dagger} [(\bH_i^u \bA)(\bH_i^u \bA)^{\dagger}+\sum_{i'=1}^{i-1}\frac{\SNR_{i'}^u}{\SNR_i^u}(\bH_i^u \bA)(\bH_i^u \bA)^{\dagger}+\frac{1}{\SNR_i^u} \bI]^{-1}$, $\bB_{ii}^u = \bC_i^u\bH_i^u\bA$ and $\bB_{ii'}^u = \bC_{i}^u\bH_{i'}^u\bA$. Thus combining \eqref{eqn:Tx_symbols_UL_OTFS} and \eqref{eqn:Rcvd_UL_OTFS}, \eqref{eqn:data_ULMMSEOTFS} can be rewritten as:   
\begin{align}
\label{eqn:MMSE_OTFS_desired_interf_UL}
\hat{\bd_i^u}= \underbrace{\bB_{ii}^u\sqrt{\power_i^u}\bd_i^u}_\text{desired signal}&+\underbrace{\sum_{i'=1}^{i-1}\bB_{ii'}^u\sqrt{\power_{i'}^u}\bd_{i'}}_\text{NOMA interference}\\ \nonumber
&+\underbrace{\bC_i^u\bn_i}_\text{noise component},~ i=1,\cdots, K,
\end{align}
Expanding \eqref{eqn:MMSE_OTFS_desired_interf_UL}, the uplink SINR for $j$-th symbol of the $i$-th user can be formulated as:
\begin{equation}
\label{eqn:SINR_UL}
\SINRrxUL_{ij} = \frac{\underbrace{\power_i^u|b^{uii}_{jj}|^2}_\text{desired power}}{\begin{aligned} \bigg[\underbrace{\power_i^u\sum_{l=1, l\neq j}^{MN}|b^{uii}_{jl}|^2}_\text{inter-symbol interference}+ &
\underbrace{\sum_{i'=1}^{i-1}\power_{i'}^u(\sum_{l=1}^{MN}|b^{uii'}_{jl}|^2)}_\text{NOMA interference}   + \\
&\underbrace{\sum_{l=1}^{MN}|c^{ui}_{jl}|^2\sigma_n^2}_\text{noise power}\bigg]\end{aligned}},
\end{equation}
with $i=1,\cdots, K$ and $j=1, \cdots, MN$. $b^{uii}_{pq}$, $b^{uii'}_{pq}$ and $c^{ui}_{pq}$ denote the $(p,q)^{\text{th}}$ elements of $\bB_{ii}^u$, $\bB_{ii'}^u$ and $\bC_i^u$ respectively. 

Using similar assumptions made for downlink direction, the sum rate (in bps/Hz) in uplink direction is given by: 
\begin{equation}
\label{eqn:UL_sum_rate_expression}
R_{sum}^{UL} = \sum_{i=1}^K \text{log}_2(1+\SINRrxUL_i). 
\end{equation}
\section{Power Allocation Schemes among Downlink NOMA-OTFS Users}
\label{sec:Pow_allocate_DL_NOMA}
 In this section, we briefly outline the various power allocation schemes used for performance evaluation of NOMA-OTFS in downlink direction. \\
 \indent It is noteworthy that in high-mobility scenarios, where OTFS offers distinguishable advantages over OFDM, feeding back full instantaneous CSI to the BS becomes increasingly difficult due to coherence time constraint of the channel. However, we also describe the full CSI based power allocation algorithm, which is used as a benchmark in comparative study. \\
 \indent It is also worth noting that in this work, while considering power allocation for weighted sum rate maximization (discussed in Sec. \ref{subsec:WSRM}), we have considered only two users multiplexed together following the principle of NOMA for analytical simplicity. Extension to generalized $K$-user scenario can be seen as a potential future investigation, which is beyond the scope of the current work. 
 \subsection{Fixed Power Allocation (FPA)}
 This power allocation scheme is simplistic in nature where the fractions of transmit power for different users are determined a priori. Such conventional scheme has been used in NOMA performance analysis for simplicity and in order to have a benchmark for other sophisticated power allocation strategies \cite{Saito_2013a, Hou_2018}. The power fractions are independent of user channel conditions and system performance. In order to maintain fairness among users, it is a general practice to allocate more power to the users with lower received average SNR. Thus, for the SNR order mentioned in \ref{subsec:Ana_DL}, the fixed transmit power fractions will be ordered as: $\NomaPowFrac_1>\NomaPowFrac_2 \cdots > \NomaPowFrac_K$, with the constraint $\sum_{i=1}^K \NomaPowFrac_i = 1$. 
 \subsection{Fractional Transmit Power Allocation (FTPA)}
 In this dynamic power allocation scheme, the fraction of power allocated to any user is proportional to the inverse of its channel gain so that the users with lower channel gain gets greater transmit power in order to maintain system fairness \cite[Sec. II.B]{Saito_2013b}. Depending on the nature of information about the channel available at the BS, following two FTPA schemes have been investigated in this work:
 \subsubsection{Average SNR based FTPA}\label{subsec:FTPA_AvgSNR}
 In this scheme, it is assumed that the BS has the access to only slowly time-varying average received SNR values ($\SNR_i$) of the users (which can be fed back to the BS from user through feedback channel or measured in reverse channel). Therefore, we propose to use the average received SNR values to allocate the users' power fractions using FTPA. The fraction of power allocated to the $i$-th user is given by:
 \begin{equation}
 \label{eq:NOMA_pow_frac_FTPA_Av_SNR}
  \NomaPowFrac_i = \frac{\SNR_i^{-1}}{\sum_{i'=1}^K \SNR_{i'}^{-1}}.
 \end{equation}
 \subsubsection{Channel Norm based FTPA}\label{subsec:FTPA_Channorm}
 The base station has access to the partial CSI of all users in terms of the instantaneous channel norms, it is assumed that we use those values to evaluate the users' power fractions. Therefore, the fraction of power allocated to the $i$-th user can be expressed as:
 \begin{equation}
 \label{eq:NOMA_pow_frac_FTPA_ChanNorm}
  \NomaPowFrac_i = \frac{||\bH_i||_F^{-1}}{\sum_{i'=1}^K ||\bH_{i'}||_F^{-1}},
 \end{equation}
 where $\bH_i$ is defined in \eqref{eq:H_chan_matrix}.
 
\subsection{Power Allocation for Weighted Sum Rate Maximization (WSRM)}\label{subsec:WSRM}
Similar to the case for FTPA, we present two weighted sum rate maximization framework based on average SNR information and instantaneous channel information at the base station. 
 \subsubsection{Average SNR based WSRM}\label{subsec:WSRM_AvgSNR}
 In case the base station has access to the average SNR information of the users, the optimization problem can be formulated based on the AWGN rates as described below:

\begin{align}
\label{eq:AWGN_rate_max_problem}
 \text{Maximize}~~~R_{sum}^{AWGN}&= w_1\text{log}(1+\frac{\NomaPowFrac_1\SNR_1}{1+\NomaPowFrac_2\SNR_2}) \\ \nonumber
 &~~~~~~~~~~~~~+w_2\text{log}(1+\NomaPowFrac_2\SNR_2)\\ \nonumber
 &  \text{subject to}~~\NomaPowFrac_1+\NomaPowFrac_2=1, 0\leq\NomaPowFrac_1, \NomaPowFrac_2\leq1;
\end{align}
where $w_1$ and $w_2$ are the weights assigned to the two users in order to maintain fairness in power allocation.
This being an early work, we obtain a suboptimal solution of the maximization problem by differentiating the cost function, as done in \cite[Sec. III.A]{Nain_2017}, although concavity of such cost function in \eqref{eq:AWGN_rate_max_problem} is not straightforward to be shown.
Reducing the problem in terms of only $\NomaPowFrac_2$, differentiating $R_{sum}^{AWGN}$ w.r.t. $\NomaPowFrac_2$ and equating it to zero finally yields:
\begin{equation}
 \label{eq:AWGN_rate_differntiation}
 \frac{w_1\SNR_2}{1+\NomaPowFrac_2\SNR_2} - \frac{w_2\SNR_1}{1+\NomaPowFrac_2\SNR_1}=0.
\end{equation}
Solving the linear equation, the optimal value of $\NomaPowFrac_2$ can be obtained as:
\begin{equation}
\label{eqn:AWGN_NOMA_opt_pow_frac}
 \NomaPowFrac_2^{Opt} = \frac{w_2\SNR_1-w_1\SNR_2}{(w_1-w2)\SNR_1\SNR_2}. 
\end{equation}
In order to impose the associated constraints stated in \eqref{eq:AWGN_rate_max_problem}, we assign $\NomaPowFrac_2^{Opt} = \text{max}(0,\text{min}(1,\NomaPowFrac_2^{Opt}))$. Clearly, $\NomaPowFrac_1^{Opt} = 1-\NomaPowFrac_2^{Opt}$. 
\subsubsection{Instantaneous Channel Information based WSRM}\label{subsec:WSRM_instant}
If the base station has access to partial information about the instantaneous channel of each user (in terms of $\bB_i$ and $\bC_i$ matrices defined after \eqref{eqn:data_DLMMSEOTFS}), the 2-user optimization problem can be formulated using the exact post-SIC SINR expression (for $j$-th symbol) derived in \eqref{eqn:PostSIC_SINR_DL} as follows:
\begin{equation}\label{eq:Exact_rate_max_problem}
\begin{split}
  \mathrm{Maximize} \  R_{sum}^{Inst}= w_1\text{log}(1+\SINRpostrxDL_{1j})+w_2\text{log}(1+\SINRpostrxDL_{2j})\\
  \text{subject to}~~\NomaPowFrac_1+\NomaPowFrac_2=1, 0\leq\NomaPowFrac_1, \NomaPowFrac_2\leq1. 
\end{split}
\end{equation}
Using the notations $\power|b^1_{jj}|^2 = \Gamma_{1d}$, 
$\power\sum_{l=1, l\neq j}^{MN}|b^1_{jl}|^2=\Gamma_{1\text{ISI}}$, 
$\power(\sum_{l=1}^{MN}|b^1_{jl}|^2) = \Gamma_{1\text{N}}$, 
$\sigma_n^2\sum_{l=1}^{MN}|c^1_{jl}|^2 = P_{1n}$,
$\power|b_{jj}^2|^2 = \Gamma_{1d}$, 
$\power\sum_{l=1, l\neq j}^{MN}|b^2_{jl}|^2=\Gamma_{2\text{ISI}}$, and
$\sigma_n^2\sum_{l=1}^{MN}|c^2_{jl}|^2 = P_{2n}$, the instantaneous weighted sum rate in terms of $\NomaPowFrac_2$ can be expressed as:
\begin{align}
\label{eqn:Inst_Rate_expression}
R_{sum}^{Inst} &= w_1\text{log}[\frac{(1-\NomaPowFrac_2)(\Gamma_{1d}+\Gamma_{1\text{ISI}})+\NomaPowFrac_2\Gamma_{1\text{N}}+P_{1n}}{(1-\NomaPowFrac_2)\Gamma_{1\text{ISI}}+\NomaPowFrac_2\Gamma_{1\text{N}}+P_{1n}}]+\\ \nonumber
&~~~~~~~~~~~~~~~~~~~~~~w_2\text{log}[\frac{\NomaPowFrac_2(\Gamma_{2d}+\Gamma_{2\text{ISI}})+P_{2n}}{\NomaPowFrac_2\Gamma_{2\text{ISI}}+P_{2n}}].
\end{align}
\begin{figure*}
\small
\begin{align}
\label{eqn:Inst_Rate_diff}
 \frac{w_1(\Gamma_{1\text{N}}-\Gamma_{1\text{ISI}}-\Gamma_{1d})}{(1-\NomaPowFrac_2)(\Gamma_{1d}+\Gamma_{1\text{ISI}})+\NomaPowFrac_2\Gamma_{1\text{N}}+P_{1n}}-
\frac{w_1(\Gamma_{1\text{N}}-\Gamma_{1\text{ISI}})}{(1-\NomaPowFrac_2)\Gamma_{1\text{ISI}}+\NomaPowFrac_2\Gamma_{1\text{N}}+P_{1n}}+\frac{w_2(\Gamma_{2d}+\Gamma_{2\text{ISI}})}{\NomaPowFrac_2(\Gamma_{2d}+\Gamma_{2\text{ISI}})+P_{2n}}-\frac{w_2\Gamma_{2\text{ISI}}}{\NomaPowFrac_2\Gamma_{2\text{ISI}}+P_{2n}}=0
\end{align} 
\rule{0.97\textwidth}{0.5pt}
\end{figure*}
As done in previous section, we use the differentiation method in order to obtain a suboptimal solution of $\NomaPowFrac_2$ \cite[ Sec. III.A]{Nain_2017}.
Differentiating $R_{sum}^{Inst}$ w.r.t. $\NomaPowFrac_2$ and equating it to zero results in \eqref{eqn:Inst_Rate_diff}. By numerically solving \eqref{eqn:Inst_Rate_diff} using available software tools,
optimal value of $\NomaPowFrac_2$ ($\NomaPowFrac_2^{Opt}$) can be obtained. Similar to the earlier case, we finally assign $\NomaPowFrac_2^{Opt} = \text{max}(0,\text{min}(1,\NomaPowFrac_2^{Opt}))$ and $\NomaPowFrac_1^{Opt} = 1-\NomaPowFrac_2^{Opt}$. \\
\indent It is to be noted that judicious assignment of weights for the users has been addressed in literature considering proportional fairness \cite{Seyama_2015, Nain_2017}. However, this being an early investigation, for simplicity we consider assignment of fixed weights as: $w_1 = 0.6, w_2 = 0.4$.\\
\indent It is also important to note that in order to implement the power allocation schemes described in Sec. \ref{subsec:FTPA_Channorm} and \ref{subsec:WSRM_instant}, channel information (like channel norm for the first scheme and $\bB$ and $\bC$ matrices for the second scheme) have to be either fed back to the BS by an error-free feedback channel, or measured at BS itself by exploiting uplink-downlink duality. Such channel measurement and feedback has to be done at least once in every delay-Doppler coherence time of the OTFS channel. It has been reported in literature that the delay-Doppler coherence time of OTFS channel is significantly larger than the coherence time in time-frequency domain for OFDM \cite{Monk_2016}. Thus, instantaneous De-Do CSI based NOMA power allocation schemes in high Doppler scenarios are easily realizable in OTFS systems compared to OFDM systems.
\section{\textcolor{black}{Link Level Performance Analysis of NOMA-OTFS Systems}}
\label{sec:link_level_OTFS_NOMA}
\indent It has been reported in literature that scheduling high number of users (more than 2-3) in power domain NOMA in same resource block does not offer much gain despite prohibitively increasing the complexity of transmit signal processing, signaling overhead as well as realizing the successive interference canceler receiver in polynomial or exponential order \cite{Wang_2018}. Therefore, we limit this early investigating work to 2 user multiplexed system as in~\cite{Saito_2015, Yan_2015, Yuan_2018}.
\begin{figure}
\hbox{\hspace{-0.5em} \includegraphics[width= \linewidth]{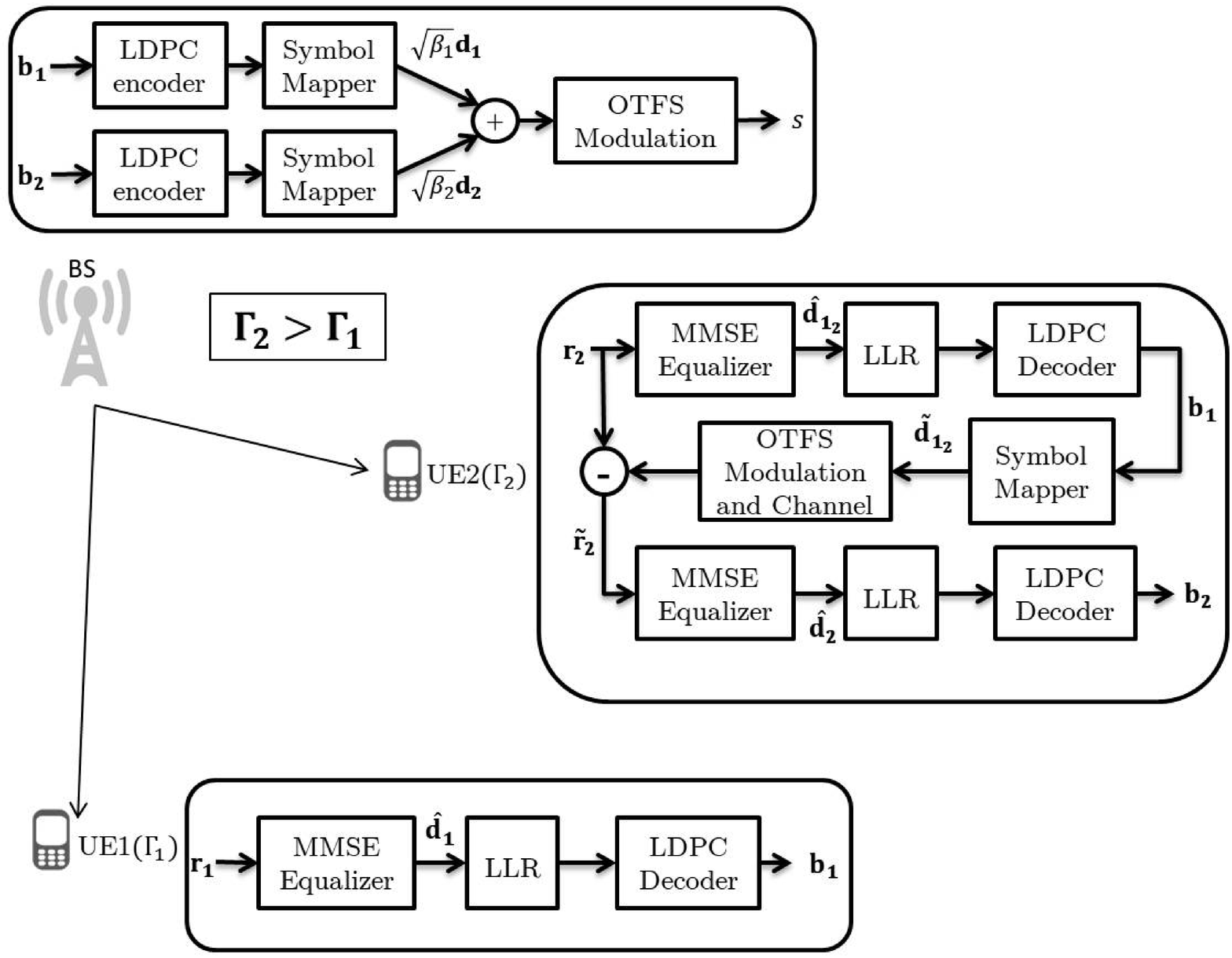}}
 \caption{\small{Representative block diagram of 2-user NOMA-OTFS system in downlink.}}
 \label{fig:DL_NOMA_SIC_BD}
\end{figure}
\subsection{Downlink MMSE SIC Receiver with LDPC coding}\label{subsec:Rx_LDPC_DL}
\indent This section illustrates the practical realization of a 2-user LDPC enabled codeword level SIC OTFS-NOMA receiver for downlink transmission as highlighted in Fig. \ref{fig:DL_NOMA_SIC_BD}. The base station generates the data for both the users (denoted as $\bb_1$ and $\bb_2$ respectively), encode using the LDPC encoder and then modulate the data using modulation supported by the user. The encoded signals for both the users are denoted as $\bd_1$ and $\bd_2$ respectively in Fig. \ref{fig:DL_NOMA_SIC_BD}.
The modulated symbols are further superimposed with allocated power ($\beta_i$). The superimposed time-domain signal is further modulated for OTFS using SFFT and the Heisenberg transform. The resulting signal $\bs$ (refer to \eqref{eqn:Superposed_Tx_symbols} for its mathematical expression) is broadcast through the delay-Doppler channel to both the users.\\
\indent Since in this work, we consider 2 users, we let $K=2$ in \eqref{eqn:Rcvd_DL_OTFS}. Both the users first perform LMMSE equalization in order to mitigate the ISI and ICI.
Additionally, as it is assumed that SNR of second user is higher than the first user, thus second user performs the SIC.
\subsubsection{Processing at First User}\label{subsubsec:1st_UE_decoding_DL}
The equalized data using the MMSE equalizer can be described using \eqref{eqn:data_DLMMSEOTFS} with $i=1$.
In order to decode the equalized data using LDPC decoder, the channel LLR values 
are calculated from the equalized symbols as,
\begin{equation}
\label{eq:LLR_bitwise_DL_UE1}
\mathcal{L}(b_{1\eta}^j | \hat{\bd}_1(\eta)) \approx ({\min_{s\epsilon S_{j}^{0}}}\frac{||\hat{\bd}_1(\eta)-s||^{2}}{\boldsymbol{\sigma_1^2}(\eta)}) - ({\min_{s\epsilon S_{j}^{1}}}\frac{||\hat{\bd}_1(\eta)-s||^{2}}{\boldsymbol{\sigma_1^2}(\eta)})
\end{equation}
where $\bd_{i}(\eta)$ is the $\eta^{th}$ element of $\bd_{i}$ mapped from the bits $b_{i\eta}^{0}~b_{i\eta}^{1}\cdots b_{i\eta}^{K_i-1}, K_i$ is the number of bits per symbol for user $i$  and $\boldsymbol{\sigma_1^2}(\eta)$ is the $\eta$th element of $\boldsymbol{\sigma_1^2} = \frac{1}{\beta_1} diag(\sigma_n^2 \bC_1 \bC_1^{\dagger} + \beta_2 \bB_1\bB_1^{\dagger})$. The aggregate interference and noise is assumed to follow Gaussian distribution, as previously stated in Sec. \ref{subsec:Ana_DL}.  $S_j^k$ denotes the set of constellation symbols in which the bit $b^j = k$. 
See the paragraph following \eqref{eqn:MMSE_OTFS_desired_interf_DL} for the definition of $\bB_1$ and $\bC_1$ matrices.
These LLRs are then fed into the LDPC decoder to decode first user's data.
Let $\bL^1$ denotes a matrix where $\bL^1(\eta,j) = \mathcal{L}(b_{1\eta}^j | \hat{\bd}_1(\eta))$ for $\eta = 1,2,\cdots,MN$ and $j=0,1,\cdots,K_i-1$. 
$\bL^1$ is reshaped to $L_{cl}\times N_{cw}$ matrix where $L_{cl}$ and $N_{cw}$ denote the LDPC codeword length and number of codewords respectively.
Each column of $\bL^1$ subsequently regenerates codeword $c^{1}_{\iota}$ for $\iota=1,2\cdots,N_{cw}$ using the Min-Sum Algorithm \cite{Zhao_2005} employed by the LDPC decoder.
This algorithm iteratively updates the variable node and check node equation as discussed below.
\begin{itemize}
\item Variable Node Update 
\begin{equation}
m_{\mu,\nu}^{(l)} = Z_{\mu} + \Sigma_{\nu' \neq \nu}m_{\mu',\nu}^{(l-1)} 
\end{equation}

where the $Z_\mu$ is the channel LLR calculated from \eqref{eq:LLR_bitwise_DL_UE1} for the $\mu$th bit in the codeword and $m_{\nu,\mu}^{(l)}$ is the message received from the $\nu$th check node to the $\mu$th variable node in the iteration $l$.
\item Check Node Update
\begin{equation}
m_{\nu,\mu}^{(l)} = \Pi_{\mu' \neq \mu} sign(m_{\mu',\nu}^{(l)})\min_{\mu' \neq \mu}(|m_{\mu',\nu}^{(l)}|)
\end{equation}
where the product and the minimum operator is taken over the set of neighboring variable nodes except the message recipient itself.
\item Decoding Decision
\begin{equation}
Z_{\mu}^{tot} = Z_{\mu} + \Sigma_{\nu}m_{\nu,\mu}^{(l)}
\end{equation}
The algorithm terminates when the termination conditions of LDPC termination are satisfied or iteration count reaches the maximum number of iterations($N_{i_{max}}$) and the decoded codeword bit $c_\mu = 1$ if $Z_{\mu}^{tot} >= 0$ and $c_\mu = 0$ if $Z_{\mu}^{tot} < 0$.
\end{itemize}
\subsubsection{Processing at Second User}\label{subsubsec:2nd_UE_decoding_DL}
\indent Since second user experiences higher SNR, it performs the SIC in which it decodes first user's data and then uses it to cancel the interference to decode its own data. The detected first user's data at the second user is given as,
\begin{equation}
 \label{eq:}
 \hat{\bd}_{1_2}=\sqrt{\beta_1}(\bH_2\bA)^{\dagger} [(\bH_2 \bA)(\bH_2 \bA)^{\dagger}+ \frac{1}{\SNR_2} \bI]^{-1} \br_2
\end{equation}
 
Corresponding LLR of the equalized data of fist user is calculated as,
\begin{equation}
\mathcal{L}(b_{{1_2}\eta}^j | \hat{\bd}_{1_2}(\eta)) \approx ({\min_{s\epsilon S_{j}^{0}}}\frac{||\hat{\bd}_{1_2}(\eta)-s||^{2}}{\boldsymbol{\sigma_{1_2}^2}(\eta)}) - ({\min_{s\epsilon S_{j}^{1}}}\frac{||\hat{\bd}_{1_2}(\eta)-s||^{2}}{\boldsymbol{\sigma_{1_2}^2}(\eta)}), 
\end{equation}
where $\boldsymbol{\sigma_{1_2}^2}(\eta)$ is the $\eta^{th}$ element of $\boldsymbol{\sigma_{1_2}^2} = \frac{1}{\beta_1} diag(\sigma_n^2 \bC_2 \bC_2^{\dagger} + \beta_2 \bB_2\bB_2^{\dagger})$.
the residual received signal at second user after canceling the interference due to first user is given by,
\begin{equation}
\label{eq:residual_signal_UE2}
\tilde{\br}_2 = \br_2 - \sqrt{\beta_1 \power}\bH_2\bA\tilde{\bd}_{1_2}, 
\end{equation}
where $\tilde{\bd}_{1_2}$ is generated at second user after passing the LDPC decoded codeword obtained from $\hat{\bd}_{1_2}$ through symbol mapper. After doing MMSE equalization on the residual signal given in \eqref{eq:residual_signal_UE2}, the detected second user's data at the second user itself is given by,
\begin{equation}
\hat{\bd}_2=\sqrt{\beta_2}(\bH_2\bA)^{\dagger} [\beta_2(\bH_2 \bA)(\bH_2 \bA)^{\dagger}+ \frac{1}{\SNR_2} \bI]^{-1} \tilde{\br}_2 
\end{equation}
As done for first user, the bit level LLRs for second user from the symbols are calculated as,
\begin{equation}
\label{eq:LLR_bitwise_DL_UE2}
\mathcal{L}(b_{2\eta}^j| \hat{\bd}_2(\eta)) \approx ({\min_{s\epsilon S_{j}^{0}}}\frac{||\hat{\bd}_2(\eta)-s||^{2}}{\boldsymbol{\sigma^2_2}(\eta)}) - ({\min_{s\epsilon S_{j}^{1}}}\frac{||\hat{\bd}_2(\eta)-s||^{2}}{\boldsymbol{\sigma^2_2}(\eta)})
\end{equation}
where $\boldsymbol{\sigma_{2}^2}(\eta)$ is the $\eta^{th}$ element of $\boldsymbol{\sigma_2^2} = \frac{1}{\beta_2} diag(\sigma_n^2 \bC_2 \bC_2^{\dagger})$ (see the paragraph following \eqref{eqn:MMSE_OTFS_desired_interf_DL} for the definition of $\bC_2$ matrix).
The LLRs are updated and the data of user 2 is generated by the LDPC decoder to generate the data using Min-Sum
algorithm as described in detail for user 1 in Sec. \ref{subsubsec:1st_UE_decoding_DL}.
\begin{figure}
\hbox{\hspace{-0.5em} \includegraphics[width= \linewidth]{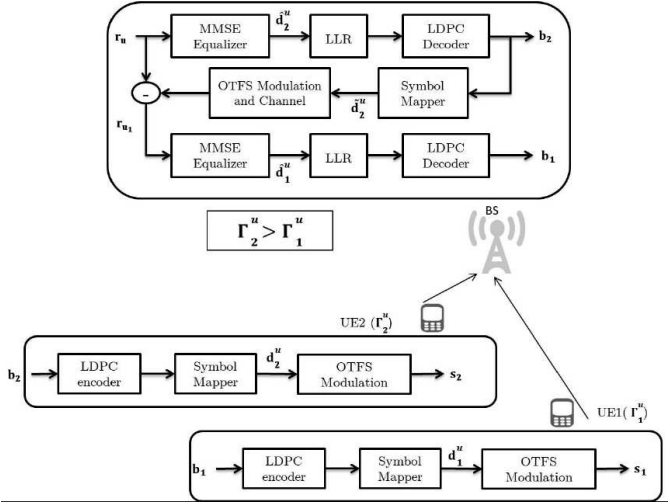}}
 \caption{\small{Representative block diagram of 2-user NOMA-OTFS system in uplink.}}
 \label{fig:UL_NOMA_SIC_BD}
 \vspace{-0.8cm}
\end{figure}
\subsection{Uplink MMSE SIC Receiver with LDPC coding}\label{subsec:Rx_LDPC_UL}
This section focuses on a realization of an OTFS-NOMA link coupled with LDPC codes for a two user scenario in uplink transmission as highlighted in Fig. \ref{fig:UL_NOMA_SIC_BD}.
For two user case, the received signal can be expressed using \eqref{eqn:Rcvd_UL_OTFS} by putting $K=2$. 
At the BS, since it is assumed $\SNR_1^u < \SNR_2^u$, we first decode second user's data as,
\begin{equation}
\hat{\bd}_2^u=(\bH_2^u\bA)^{\dagger} [(\bH_2^u \bA)(\bH_2^u \bA)^{\dagger} + \frac{\SNR_1^u}{\SNR_2^u}(\bH_1^u \bA)(\bH_1^u \bA)^{\dagger} + \frac{1}{\SNR_2^u} \bI]^{-1} \br_u. 
\end{equation}
LLR values of second user can be computed as,
\begin{equation}
\mathcal{L}(b_{2\eta}^{u^j} | \hat{\bd}_2^u(\eta)) \approx ({\min_{s\epsilon S_{j}^{0}}}\frac{||\hat{\bd}^u_2(\eta)-s||^{2}}{\boldsymbol{\sigma^2_2}(\eta)}) - ({\min_{s\epsilon S_{j}^{1}}}\frac{||\hat{\bd}^u_2(\eta)-s||^{2}}{\boldsymbol{\sigma^2_2}(\eta)}), 
\end{equation}
where $\eta^{th}$ element of  $\bd^u_{i}$, $\bd^u_{i}(\eta)$ is mapped from bits $b_{i\eta}^{u^0}~b_{i\eta}^{u^1}\cdots b_{i\eta}^{u^{(K_i-1)}}, K_i$ is the number of bits per symbol for user $i$ and $\boldsymbol{\sigma^2_2}(\eta)$ is the $\eta^{th}$ element of $\boldsymbol{\sigma_2^2} = \frac{1}{\power_2} diag(\sigma_n^2 \bC_2^u \bC_2^{u\dagger} ~ + ~ \power_1\bB_{21}^u\bB_{21}^{u\dagger})$. The matrices $\bC_2^u$ and $\bB_{21}^u$ are defined after \eqref{eqn:data_ULMMSEOTFS}.
The calculated LLR values are processed by LDPC decoder in order to produces the message word. The obtained message is again encoded and modulated to generate the recovered data $\tilde{\bd}^{u}_2$ for second user, which is used to cancel the interference from aggregate received signal to decode the first user's data as,
\begin{equation}
\label{eqn:residual_signal_UE1}
\tilde{\br}_{u_1} = \br_u - \sqrt{\power_2^u}\bH_2^u\bA\tilde{\bd}^u_2.
\end{equation}
After doing MMSE equalization of residual signal at the BS given by \eqref{eqn:residual_signal_UE1}, the detected first user's data is given by,
\begin{equation}
 \hat{\bd}^u_1=(\bH_1\bA)^{\dagger} [(\bH_1 \bA)(\bH_1 \bA)^{\dagger}+ \frac{1}{\SNR_1} \bI]^{-1} \tilde{\br}_{u_1}.
\end{equation}
The equalized data is $\hat{\bd}^u_1$ is used to calculate the LLR as follows:
\begin{equation}
\mathcal{L}(b_{1\eta}^j | \hat{\bd}^u_1(\eta)) \approx ({\min_{s\epsilon S_{j}^{0}}}\frac{||\hat{\bd}^u_1(\eta)-s||^{2}}{\boldsymbol{\sigma^2_1}(\eta)}) - ({\min_{s\epsilon S_{j}^{1}}}\frac{||\hat{\bd}^u_1(\eta)-s||^{2}}{\boldsymbol{\sigma^2_1}(\eta)}), 
\end{equation}
where assuming perfect SIC, $\hat{\bd}^u_1(\eta)$ and $\boldsymbol{\sigma^2_2}(\eta)$ are the $\eta^{th}$ element of $\hat{\bd}$ and $\boldsymbol{\sigma_1^2} = \frac{1}{\power_1^u} diag(\sigma_n^2 \bC_1^u \bC_1^{u\dagger})$. The matrix $\bC_1^u$ is also defined after \eqref{eqn:data_ULMMSEOTFS}.
The calculated LLR values are then fed to LDPC decoder to reproduce the data of user 1.
\begin{table}[h]
\caption{\small{Key system parameters}}
\centering
\begin{tabular}{|m{4.5cm}|m{3cm}|}\hline
\textbf{Parameter} & \textbf{Value}\\ \hline
LTV Delay-Doppler channel model & `Extended Vehicular A (EVA)'\cite{itu2135} \\ \hline
Doppler slots ($N$) & $16$ \\ \hline
Delay slots ($M$) & $256$  \\ \hline
Number of NOMA users & 2  \\ \hline
User speed &  $500$ kmph \\ \hline
Carrier frequency & $5.9$ GHz \\ \hline
Subcarrier Bandwidth $\Delta f$ & $15$ KHz \\ \hline
Total Bandwidth $B$ & $3.84$ MHz \\ \hline
Frame Duration $T_f$ & $1.08$ ms \\ \hline
Error Correction Codes & LDPC codes. Code length = 648,code rate (R) = 2/3 \cite{IEEE_80211n_standard}\\ \hline
Downlink average SNR & $\SNR_1 = 15$ dB, $\SNR_2 = 25$ dB\\ \hline
Uplink average SNR & $\SNR_1^u = 10$ dB, $\SNR_2^u = 30$ dB\\ \hline
\end{tabular}
\label{table:SystemParameter}
\end{table}
\section{Simulation Results and Discussion}\label{sec:result_discussion}
In this section we present detailed performance analysis in terms of system level and link level evaluation of the presented NOMA-OTFS schemes in high speed scenario through extensive Monte-Carlo simulation for both downlink and uplink. 
The important simulation parameters are listed in Table~\ref{table:SystemParameter}. Doppler is generated using Jake's formula, $\nu_p=\nu_{max} cos(\theta_p)$, where $\theta_p$ is uniformly distributed over $[-\pi ~ \pi]$. The CP is chosen long enough to accommodate the maximum excess delay of the channel. 
We present results of an equivalent OFDM system with synchronous CP length and block based signal structure as described in Sec. \ref{sec:OTFS_basic}.
The MMSE equalizer implemented in this work can efficiently cancel the ICI at the receiver, unlike the single tap equalizer used in traditional systems \cite{Jeon_1999}.  \\
\indent While 5G-NR has provision for variable subcarrier bandwidth \cite{Das_2008, Zaidi_2016}, the EVA channel model restricts us to a subcarrier bandwidth of upto 60 KHz corresponding to numerology 2 contrary to the maximum of 240 KHz with numerology 4 due to the coherence bandwidth of about 56 KHz. 
However, it is worth mentioning that from ICI perspective, numerology 4 with maximum of 240 KHz is more desirable.
It is important to note that the receiver used here is MMSE with ideal channel estimates thus serving the purpose to cancel inter sub-carrier interference of LTV channel which appear due to Doppler spread. 
The system performance is then accordingly evaluated using a subcarrier bandwidth of 15 KHz which is also valid for system design of 4G systems \cite{Sesia_2011, Dahlman_2013}. \\
\indent In Sec. \ref{subsec:SE_results}, we present the system level performance of downlink and uplink in terms of sum SE results for the MMSE-SIC receiver developed in \ref{subsec:Ana_DL} and \ref{subsec:Ana_UL} respectively. 
In Sec. \ref{subsubsec:SE_OMA_NOMA_results}, we evaluate the performance results of NOMA-OTFS and OMA-OTFS in order to find the feasibility of NOMA-based multi-user multiplexing in high mobility scenarios.  
We then conduct a comparative performance analysis between OTFS based and OFDM based NOMA implementation with an aim to explore relative gains that can be obtained by such system design in high Doppler scenarios (in Sec. \ref{subsubsec:SE_OTFS_OFDM_results}). Since we evaluate different power allocation for NOMA-OTFS in Sec. \ref{sec:Pow_allocate_DL_NOMA}, we analyze their relative performance in Sec. \ref{subsubsec:SE_Pow_Alloc_results}. 
After system-level performance evaluation, we delve into link level performance verification of LDPC coded NOMA-OTFS CWIC MMSE-SIC receiver in Sec. \ref{subsec:BLER_OTFS_OFDM_results}. We also compare the same against the equivalent NOMA-OFDM as well as OMA-OTFS systems as described earlier in this work.
The performance is evaluated in terms of metrics like BLER ($P_{e}$), throughput and goodput. 
The throughput of a link is defined as the number of bits transmitted per unit time and is given by \eqref{eqn:throughput}.
\begin{equation} \label{eqn:throughput}
 \text{Throughput} ~ = ~ \Sigma_{i=1}^{2}\text{R}\text{K}_i  ~ \text{bits/s/Hz,}
\end{equation}
where $\text{R}$ and $\text{K}_i$ denote code rate and bits per QAM symbol respectively.
Whereas, the goodput of a link is defined as the number of bits that are successfully received and expressed in \eqref{eqn:Goodput}.
\begin{equation} \label{eqn:Goodput}
 \text{Goodput} ~ = ~ \Sigma_{i=1}^{2}\text{R}\text{K}_i(1-P_{e_i}) ~ \text{bits/s/Hz,}
\end{equation}
where $P_{e_i}$ denotes BLER for $i$-th user respectively.
\begin{figure}
\includegraphics[width= \linewidth]{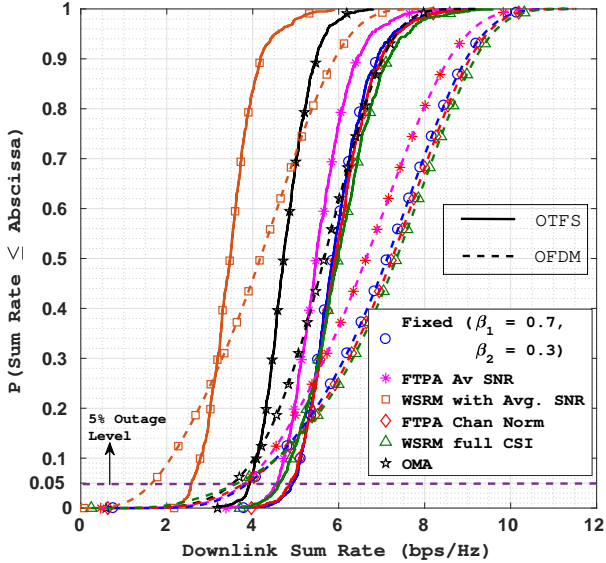}
 \caption{\small{CDF of downlink sum rate for various NOMA power allocation schemes under OTFS/OFDM for $\SNR_1=15$ dB, $\SNR_2=25$ dB for user velocity = 500 kmph. Solid and dashed lines represent OTFS and OFDM results respectively. Markers denote results corresponding to various power allocation schemes described in Sec. \ref{sec:Pow_allocate_DL_NOMA}}}
 \label{fig:cdfplot_DL_NOMA_OMA_OTFS_OFDM_15_25}
\end{figure}
 \subsection{System Level Spectral Efficiency Results}\label{subsec:SE_results}
\subsubsection{Comparison between NOMA/OMA-OTFS} \label{subsubsec:SE_OMA_NOMA_results}
In Fig. \ref{fig:cdfplot_DL_NOMA_OMA_OTFS_OFDM_15_25}, the cumulative distribution functions (CDF) of downlink sum rates achieved under various OMA and various NOMA power allocation schemes for OTFS and OFDM are shown. The exact values of mean and $5\%$ sum rate are given in Table \ref{table:DL_mean_outage_SE_OTFS_OFDM_15_25}. From the CDF curves and tabulated values we observe a significant increase in mean and outage sum rate in the NOMA-OTFS scheme compared to the OMA-OTFS. We notice that there is more than 16\% improvement in both mean and 5\% outage sum SE respectively in case of NOMA-OTFS with average SNR based FTPA power allocation with respect to OMA-OTFS. The gain is even higher for power allocation schemes like channel-norm based FTPA and instantaneous CSI based WSRM NOMA-OTFS schemes. 
\subsubsection{Comparison between OTFS and OFDM performances} \label{subsubsec:SE_OTFS_OFDM_results}
\begin{figure}[h]
   \centering
\includegraphics[width= \linewidth]{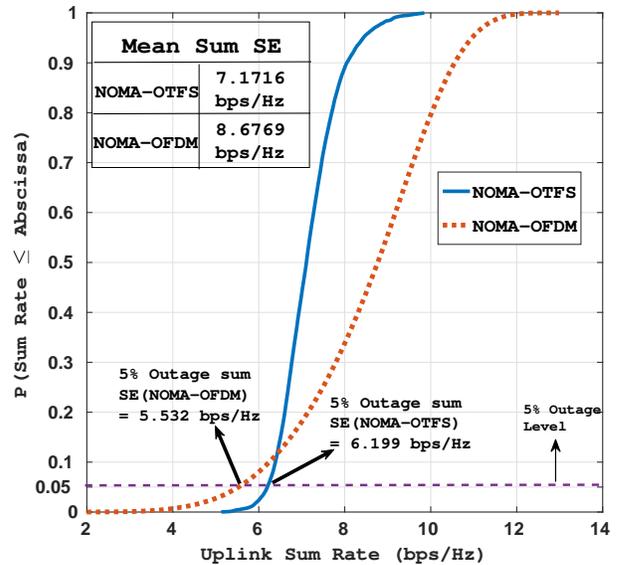}
\caption{\small{CDF of uplink sum rate for NOMA users under OTFS/OFDM for $\SNR_1^u=10$ dB, $\SNR_2^u=30$ dB for user velocity = 500 kmph.}}
   \label{fig:cdfplot_UL_NOMA_OMA_OTFS_OFDM}
\end{figure}

From the CDF trends in Fig. \ref{fig:cdfplot_DL_NOMA_OMA_OTFS_OFDM_15_25} and tabulated values in Table \ref{table:DL_mean_outage_SE_OTFS_OFDM_15_25}, it can be observed that the $5\%$ outage sum SE shows significant improvement for OTFS with respect to OFDM. For example, in case of average SNR based FTPA and channel norm based FTPA schemes, an improvement of around 19.7\% and 27.5\% respectively is observed. The gain is even higher for weighted sum rate maximization schemes, reaching to nearly 26\% and 55\% respectively. Similar improvement (around 10\%) has been observed for OMA scheme as well, highlighting the utility of OTFS over OFDM even for orthogonal multi-user scenario. Exact values of gains are tabulated in last column of Table \ref{table:DL_mean_outage_SE_OTFS_OFDM_15_25}. The outage improvement in OTFS over OFDM reflects the diversity gain of OTFS with respect to OFDM.  
However, we note that OFDM based NOMA provides nominally higher mean SE (in the order of 5-14\%) than OTFS based NOMA. Detailed analytical treatment in order to investigate this issue is a potential future work.\\
\indent Similar improvement in 5\% outage sum rate ($\sim$ 12\%) and reduction in mean sum rate ($\sim$ 17\%) in OTFS compared to OFDM has also been reflected for uplink NOMA in Fig. \ref{fig:cdfplot_UL_NOMA_OMA_OTFS_OFDM} for the MMSE-SIC receiver designed in Sec. \ref{subsec:Ana_UL}.  
\subsubsection{Comparison of various NOMA power allocation schemes}\label{subsubsec:SE_Pow_Alloc_results}
The CDF curves in Fig. \ref{fig:cdfplot_DL_NOMA_OMA_OTFS_OFDM_15_25} and tabulated values in Table. \ref{table:DL_mean_outage_SE_OTFS_OFDM_15_25} of downlink sum rates for various NOMA power allocation schemes reveal important assessment of the schemes in the considered system model. The average SNR based weighted sum rate maximization scheme's (described in Sec. \ref{subsec:WSRM_AvgSNR}) performance is seen to be the worst compared to the other schemes. This is mainly due to the fact that strong user's average received SNR ($\SNR_2$) is significantly higher than the weak user's average received SNR ($\SNR_1$), thus resulting in allocation of full power to the weak user (see \eqref{eqn:AWGN_NOMA_opt_pow_frac}), effectively turning the scheme to OMA with only weak user. Judicious choice of weights ($w_i$) incorporating proportional fairness can be used to alleviate the issue. The average SNR based FTPA scheme (described in Sec. \ref{subsec:FTPA_AvgSNR}) also gets partially affected due to the same issue and thus the scheme marginally outperforms the OMA scheme. The channel norm based FTPA (described in Sec. \ref{subsec:FTPA_Channorm}) and instantaneous CSI based weighted sum rate maximization (described in Sec. \ref{subsec:WSRM_instant}) schemes have all achieved better performance in terms of mean and outage spectral efficiency which is significantly higher than the OMA scheme, which advocates for use of De-Do PD-NOMA for better system performance. It is worth noting that the NOMA power allocation methods discussed here are suboptimal. The optimistic  results obtained here promotes further investigation of such schemes which can be implemented in high Doppler scenario.
\begin{table*}
\caption{\small{Mean and $5\%$ outage SE (in bps/Hz) for downlink NOMA in OTFS and OFDM for $\SNR_1=15$ dB, $\SNR_2 = 25$ dB for user velocity = 500 kmph.}}
\centering
\begin{tabular}{|c|c|c|c|c|c|c|}\hline
\multirow{2}{*}{\begin{tabular}[c]{@{}l@{}}NOMA Power \newline Allocation Schemes \\ \\ \end{tabular}} & \multicolumn{3}{c|}{Mean SE}           & \multicolumn{3}{c|}{5\% Outage SE}           \\ \cline{2-7}
                                                                                             & OTFS & OFDM & \% gain & OTFS & OFDM & \% gain \\ \hline
OMA                                                                                          & 4.7618  & 5.5852   & -14.74\%           & 3.931  & 3.544   & 10.92\%     \\ \hline    
Fixed-I ($\NomaPowFrac_1=0.7$, $\NomaPowFrac_2=0.3$)                                                                                       & 5.9499  & 6.2898   & -5.40\%         & 4.925  & 3.9   & 26.28\%         \\ \hline
Fixed-II ($\NomaPowFrac_1=0.9$, $\NomaPowFrac_2=0.1$)                                                                                        & 5.546  & 6.5398   & -15.19\%         & 4.5  & 3.8   & 18.42\%         \\ \hline
FTPA (Avg SNR)                                                                               & 5.5487  & 6.1500   & -9.77\%         & 4.574  & 3.821   & 19.70\%        \\ \hline
WSRM (Avg SNR)                                                                           & 3.496  & 4.0838   & -14.39\%         & 2.574  & 1.658   & 55.24\%         \\ \hline
FTPA (Channel Norm)                                                                          & 5.9977  & 6.3075   & -4.91\%         & 4.874  & 3.823   & 27.46\%        \\ \hline
WSRM (Full CSI)                                                                        & 6.0254  & 6.2922   & -4.24\%         & 4.617  & 3.654   & 26.35\%         \\ \hline
\end{tabular}
\label{table:DL_mean_outage_SE_OTFS_OFDM_15_25}
\end{table*}
\begin{table*}
\caption{\small{Userwise BLER results for downlink NOMA in OTFS and OFDM. ($\beta_1 = 0.9$ and $\beta_2 = 0.1$, code rate= 2/3,UE1 using QPSK Modulation with SNR 15 dB resulting SINR  8.35 dB)}}
\begin{tabular}{| c | c | c | c | c | c | c | c | c | c | c | c |}
\hline
\multicolumn{1}{|c|}{SNR(dB)} & \multicolumn{1}{|c|}{SINR(dB)} & \multicolumn{2}{c |}{UE2 Modulation}  & \multicolumn{2}{c|}{BLER UE2} & \multicolumn{2}{c|}{BLER UE1} & \multicolumn{3}{c |}{Goodput(bits/s/Hz)} \\  
\hline
UE2 & UE2 &  OTFS & OFDM & OTFS & OFDM & OTFS & OFDM & OTFS & OFDM & \% gain \\
\hline 
22 & 12  &  QPSK(2) & QPSK(2) &  4.7x$10^{-2}$ & 1.2x$10^{-1}$ & 3x$10^{-3}$	& 1x$10^{-3}$ & 2.6 & 2.51 & 3.46 \\
 \hline
25 & 15  &  QPSK(2) & QPSK(2) &  0 & 1.3x$10^{-3}$ & 1x$10^{-3}$	& 3x$10^{-3}$ &  2.67 & 2.66  & 0.34 \\
 \hline
30 & 20  &  16QAM(4) & 16QAM(4) & 2x$10^{-3}$	& 2.2x$10^{-2}$ & 0 & 6.5x$10^{-3}$	& 3.99 & 3.93 & 1.50 \\
 \hline
35 & 25  &  64QAM(6) & 16QAM(4)	& 5.6x$10^{-2}$ & 3x$10^{-4}$ & 5x$10^{-3}$ & 5x$10^{-3}$ & 5.10 & 3.99 & 21.76\\
\hline
\end{tabular}\\ \\
\label{table:DL_BLER_OTFS}
\vspace{-0.6cm}
\end{table*}

\begin{table*}
\caption{\small{Userwise BLER results for uplink NOMA in OTFS and OFDM. (Code rate= 2/3,UE1 using QPSK Modulation with SNR 10 dB resulting SINR  10 dB)}}
\begin{tabular}{| c | c | c | c | c | c | c | c | c | c | c | c |}
\hline
\multicolumn{1}{|c|}{SNR(dB)} & \multicolumn{1}{|c|}{SINR(dB)} & \multicolumn{2}{c |}{UE2 Modulation}  & \multicolumn{2}{c|}{BLER UE2} & \multicolumn{2}{c|}{BLER UE1} & \multicolumn{3}{c |}{Goodput(bits/s/Hz)}  \\
\hline
UE2 & UE2 &  OTFS & OFDM & OTFS & OFDM & OTFS & OFDM & OTFS & OFDM & \% gain \\
\hline 
25 & 15  &  QPSK(2) & QPSK(2) &  0 & 1.3x$10^{-3}$ & 7x$10^{-2}$	& 1.8x$10^{-1}$ & 2.57 & 2.42 & 5.84 \\
\hline
30 & 20  &  16QAM(4) & 16QAM(4)	& 1x$10^{-3}$	& 2.8x$10^{-2}$ & 6.7x$10^{-2}$ & 3x$10^{-1}$	& 3.91 & 3.52 & 9.97 \\
\hline
40 & 30  &  64QAM(6) & 64QAM(6)	& 3.4x$10^{-3}$ & 3.6x$10^{-2}$ & 9.7x$10^{-2}$ & 4.1x$10^{-1}$ & 5.19 & 4.64 & 10.60 \\
\hline
\end{tabular}\\ \\
\label{table:UL_BLER_OTFS}
\vspace{-0.8cm}
\end{table*}
\subsection{Link Level Performance of NOMA-OTFS}\label{subsec:BLER_OTFS_OFDM_results}
In this section, we discuss about the link-level performance of LDPC coded codeword level SIC NOMA-OTFS system and compare it against OMA-OTFS and NOMA-OFDM system. We first discuss about the downlink NOMA performance and then the uplink. Tables \ref{table:DL_BLER_OTFS} and \ref{table:UL_BLER_OTFS} show the goodput  performance along with BLER values of the users for downlink and uplink respectively. While generating such results, the average SNR values are considered for choosing the modulation scheme in order to guarantee that the experienced BLER remains below the threshold  $10^{-1}$ \cite{Saito_2015}. 
\subsubsection{Performance of NOMA-OTFS in Downlink}
In the downlink direction, at base station, each user's data is encoded using LDPC with code rate $\text{R}=2/3$. The encoded bit stream is modulated using QPSK, 16QAM or 64QAM ($\text{K}_i = 2, 4$ and $6$ respectively). 
In order to achieve BLER less than 0.1 with LDPC code rate and length shown in Table \ref{table:SystemParameter}, SNR thresholds to support QPSK,16QAM and 64QAM for OTFS are 9.5 dB, 15 dB and 23.5 dB respectively. For OFDM, the thresholds are 10.8 dB, 18 dB and 26 dB respectively. The modulation schemes for both the users are chosen based on the average SINR($\tilde{\Upsilon}$) experienced by the users, which are functions of $\SNR_1, \SNR_2, \NomaPowFrac_1$ and $\NomaPowFrac_2$.   
In downlink, the average SINR of user 1 are obtained assuming interference as Gaussian noise is given by
\begin{equation}
\label{eq:Avg_SINR_UE1}
\tilde{\Upsilon}_1 (\text{in dB})= 10\text{log}_{10}(\frac{\beta_1\SNR_1}{\beta_2\SNR_1 + 1}).
\end{equation}
The post SIC average SINR for user 2 assuming perfect SIC can be expressed as:
\begin{equation}
\label{eq:Avg_SINR_UE2}
\tilde{\Upsilon}_2 (\text{in dB})= 10\text{log}_{10}(\beta_2\SNR_2).
\end{equation}
The modulated data of the users is transmitted using superposition coding with $\beta_1 = 0.9$ and $\beta_2 = 0.1$ as described in Sec. \ref{subsec:NOMA_OTFS_DL_SigMod}. This choice of $\beta_i$s results in $\tilde{\Upsilon}_1=8.35$ dB for $\SNR_1 = 15$ dB which puts the system close to minimum operational range. Though the SINR of 8.35 dB is insufficient to satisfy the BLER threshold as per previous discussion but it is observed that the system is able to support required BLER with these $\beta_i$s. This observation suggests that the Gaussian assumption considered for evaluating SINR may not hold true.
User 1(weak user) decodes the signal using MMSE equalization as outlined in Sec \ref{subsubsec:1st_UE_decoding_DL}. On the other hand, as detailed in Sec. \ref{subsubsec:2nd_UE_decoding_DL}, user 2(strong user) experience higher SNR and thus perform SIC at codeword level.
 Same SINR thresholds are used for uplink modulation scheme selection as in downlink. BLERs are evaluated using Monte-Carlo link level simulation for each user for downlink as well as uplink.\\
\indent Based on the obtained BLER results, we compute throughput \eqref{eqn:throughput} and goodput \eqref{eqn:Goodput} for each user. 
 Table \ref{table:DL_BLER_OTFS} is generated keeping user 1's modulation as QPSK, as $\SNR_1 = 15$ dB. $\SNR_2$ is varied such that the higher modulation schemes can be supported by user 2. $\SNR_2 = 22$, 30, 35 dB are considered. Corresponding to these $\SNR_2$ values, $\tilde{\Upsilon}_2$ obtained from \eqref{eq:Avg_SINR_UE2} support modulation schemes QPSK,16-QAM and 64-QAM respectively following the discussion made earlier about the SNR thresholds corresponding to the modulation schemes. Though the SNR range for both users can be between -3 dB and 40 dB, only these combinations of SNRs are selected as representative values in order to demonstrate system performance. An important aspect of NOMA name user selection is dependent on achievable NOMA gain, which in turn depends on supportable data rate. Here, such consideration are made from link level perspective. In operational system, EESM \cite{Donthi_2011, Song_2011} based mapping of user's experienced SINR can be done to choose appropriate rate while satisfying required BLER. \\
\indent For the SNR pair, $\SNR_1 = 15$dB and $\SNR_2 = 35$dB in the table \ref{table:DL_BLER_OTFS}, user 1 and user 2 are assigned QPSK (K=2) and 64-QAM (K=6) respectively resulting in a throughput of 5.33 bits/sec/Hz (which is evaluated from \eqref{eqn:throughput}) for OTFS while the goodput achieved is 5.10 bits/sec/Hz, which is evaluated by taking $K_1 = 2, K_2 = 6, P_{e_1} = 5.6 \times 10^{-2}, P_{e_2} = 5\times 10^{-3}$ and $R = 2/3$ in \eqref{eqn:Goodput}. For the same scenario in OMA case, user 1 can support upto 16QAM while user 2 can support upto 64QAM resulting in throughput of (4*2/3 + 6*2/3)/2 = 3.33 bits/sec/Hz.
\textit{Here the percentage gain in throughput with NOMA-OTFS over OMA-OTFS is 37.52\%}. \\
\indent When NOMA-OFDM is employed for the same conditions, user 1 is assigned QPSK and user 2 is assigned 16QAM in order to satisfy BLER threshold resulting in a goodput of 3.99 bits/s/Hz. \textit{Thus, NOMA-OTFS offers 21.76\% gain in goodput over NOMA-OFDM.}
\subsubsection{Performance of NOMA-OTFS in Uplink}
In uplink direction, the table \ref{table:UL_BLER_OTFS} is generated by keeping $\SNR_1^u = 10$ dB and varying the user 2's SNR, $\SNR_2^u = 25,30,40$ dB thus varying the user 2's modulation scheme as QPSK,16QAM and 64QAM respectively. It can be observed that NOMA using OFDM is unable to support user 1 as $P_{e_1}$ is above threshold. For the SNR pair, $\SNR_1^u = 10$dB and $\SNR_2^u = 40$dB, 
$P_{e_2} = 3.6 \times 10^{-2}$, thus user 2 can be supported with modulation scheme 64QAM but user 1 is unable to transmit even using QPSK due to the resulting BLER of about $4.1\times 10^{-1}$, as a result of error propagation. The resulting NOMA-OFDM goodput is 4.64 bits/s/Hz compared to NOMA-OTFS goodput of 5.19 bits/s/Hz, \textit{thus a gain of 10.60\% is shown for NOMA with OTFS over OFDM in uplink}. If OMA is employed for the same scenario, then user 1 and user 2 can support upto QPSK and 64QAM respectively resulting in a throughput of 2.67 bits/s/Hz, while NOMA throughput is 5.33 bits/s/Hz. \textit{Thus a gain of 50\% in throughput can be obtained in NOMA with respect to OMA.}\\
\vspace{-0.5cm}
\section{Conclusion}
In this paper, we have presented the performance analysis of a superposition coding based De-Do domain PD-NOMA-OTFS system 
in high mobility scenarios. In order to realize NOMA-OTFS, we have presented a linear MMSE-SIC receiver. Symbol-wise post-processing SINR is derived for both downlink and uplink for subsequent SE analysis of such a system. We have realized a few partial CSI-based power allocation techniques among downlink NOMA users. In order to investigate the practical applicability of such a system, we have also develop a CWIC receiver with LDPC error-correcting codes along with MMSE equalization for 2-user NOMA case.\\
\indent Results show that the De-Do domain two-user PD-NOMA-OTFS, as presented in this work, is better than traditional OMA-OTFS by upto 16\% in terms of both mean and outage sum SE performance. \\
\indent We have also observed that NOMA-OTFS has upto 50\% better outage sum SE when compared to NOMA-OFDM for partial-CSI based power allocation schemes. For full-CSI based power allocation schemes, the gain is in the order of 27\%. This also indicates that the OTFS gain over OFDM is not reduced by using NOMA.
However we note that mean sum SE of appropriately modified NOMA-OFDM is better than NOMA-OTFS. Thus we find that there is a tradeoff between mean and outage SE. The improved outage sum SE indicates a more resilient system in high-mobility scenario, which is highly desirable. \\
\indent The link level performance obtained from the developed codeword level SIC receiver shows that the NOMA-OTFS system has upto 21.76\% and 10.60\% improved goodput in downlink and uplink respectively compared to NOMA-OFDM. It also shows 37.52\% and 50\% better throughput for NOMA-OTFS over OMA-OTFS system in downlink and uplink respectively.\\
\indent Therefore, based on the system developed and presented performance analysis we find that NOMA-OTFS has the potential to improve the performance of regular OMA-OTFS and NOMA-OFDM in high mobility conditions. 

\bibliographystyle{IEEEtran}
\bibliography{OTFS_NOMA}
\end{document}